\documentclass[pdflatex,sn-mathphys-num,iicol]{sn-jnl}

\usepackage{graphicx}%
\usepackage{multirow}%
\usepackage{amsmath,amssymb,amsfonts}%
\usepackage{amsthm}%
\usepackage{mathrsfs}%
\usepackage[title]{appendix}%
\usepackage{xcolor}%
\usepackage{textcomp}%
\usepackage{manyfoot}%
\usepackage{booktabs}%
\usepackage{algorithm}%
\usepackage{algorithmicx}%
\usepackage{algpseudocode}%
\usepackage{listings}%

\newcommand{\ii}{\mathrm{i}} 

\newcommand{\tomega}{\tilde{\omega}}
\newcommand{\tphi}{\tilde{\phi}}

\newcommand{\imp}{ \mathrm{imp} }
\newcommand{\SC}{ \mathrm{SC} }
\newcommand{\kB}{ k_{\mathrm{B}} }
\newcommand{\Den}{ \mathrm{D} }
\newcommand{\CEP}{ \mathrm{CEP} }

\begin{document}

\title[Critical point of the transition between $s_\pm$ and $s_{++}$ states]{Critical point of the transition between $s_\pm$ and $s_{++}$ states of a two-band superconductor with nonmagnetic impurities}


\author*[1]{\fnm{Vadim A.} \sur{Shestakov}}\email{v\_shestakov@iph.krasn.ru}
\author[1]{\fnm{Maxim M.} \sur{Korshunov}}

\affil*[1]{\orgname{Kirensky Institute of Physics, Federal Research Center KSC SB RAS}, \orgaddress{\city{Krasnoyarsk}, \postcode{660036}, \country{Russia}}}

\abstract{Behavior of the Grand thermodynamic potential along with its derivatives, entropy and specific heat, is considered within a two-band model of an unconventional $s_\pm$ superconductor with nonmagnetic impurities. The transition $s_\pm \to s_{++}$ is shown to be a smooth crossover at high temperatures, while it becomes the first order phase transition at low temperatures. Thus, on a phase diagram `temperature'-`impurity scattering rate' there appears to be a critical end point. Temperature at which the behavior of the transition is changed is maximal in the Born limit and tends to zero away from the limit, which points out to the possible realization of a quantum phase transition.}

\maketitle

\section{Introduction.}\label{txt:intro}
Multiband superconductors such as iron pnictides and chalcogenides reveal a whole spectrum of phenomena due to the presence of many bands and interaction between them~\cite{SadovskiiReview, Hirschfeld2011}. Along such phenomenon there are $s_\pm$ unconventional superconducting state arising from spin fluctuations and a nematic state, which origin is connected to the orbital order in the multi-orbital system as well as to the spin fluctuations~\cite{Scherer2017, SteffensenKreisel2021}. With adding another ``degree of freedom'', disorder, to the multiband system, a possibility of changing the superconducting order parameter structure by controlling impurity concentration or strength of the scattering potential occurs~\cite{Hirschfeld2016, Korshunov2016}. For example, nonmagnetic impurities can induce a transition from the sign-changing $s_\pm$ order parameter to a sign-preserving $s_{++}$ one, while magnetic impurities, on the contrary, can induce a transition from the $s_{++}$ type to the $s_\pm$ one ~\cite{Korshunov2016}. Such transitions take place even within a simple two-band model with the isotropic superconducting order parameter within each band ~\cite{EfremovKorshunov2011, Dolgov2014}. Previously, $s_\pm \to s_{++}$ transition was shown to be dependent not only on the impurity scattering rate but on temperature~\cite{Koshelev2012, Silaev2017, ShestakovKorshunovSymmetry2018}. Furthermore, there is a region in the vicinity of the transition where several solutions of the Eliashberg equations coexist~\cite{ShestakovKorshunovSUST2025, ShestakovKorshunov2025FTT}.

To analyze thermodynamics of the transition, we consider the Grand thermodynamic potential $\Omega$ also known as the Landau free energy. Firstly, based on a principle of the minimum energy, we use it to choose one of the solutions of the Eliashberg equations. Secondly, the first and second derivatives of $\Omega$ with respect to temperature give us entropy and electronic specific heat, respectively. Providing that $\Omega$ is a smooth function of disorder close to the $s_\pm \to s_{++}$ transition, the transition being a crossover. However, if it has a peculiarity, kink, then entropy has a discontinuous jump, and specific heat is diverged. Such a cascade, `kink-jump-divergence', corresponds to the first order phase transition. Next, we show that for the temperature below some critical value, the crossover is changed to the first order phase transition. This temperature, in turn, depends on a `strength' of impurity scattering and has its maximum in the case of a weak scattering impurity, i.e. in the Born limit. Away from the Born limit, the critical temperature decreases. Due to the restrictions of the Matsubara technique employed, we unable to achieve the zero temperature, however an extrapolation shows that the critical temperature would tend to zero implying the existence of a quantum phase transition.

\section{The Grand thermodynamic potential.}\label{txt:omega}
It is convenient to consider change in the Grand thermodynamic potential due to the transition to superconducting state from the normal one, i.~e. the difference between Landau free energies at these states,
\begin{equation}\label{eq:DeltaOmega}
\Delta\Omega(T) = \Omega_{\mathrm{S}}(T)-\Omega_{\mathrm{N}}(T).
\end{equation}
The explicit expression for this difference was previously obtained based on the multiband generalization of the Luttinger-Ward expression, within a two-band model with $\xi$-integrated Green’s functions $\hat{\mathbf{g}}(\ii\omega_n)$ written in a combined space: band space and Nambu space~\cite{ShestakovKorshunovSUST2025, ShestakovKorshunov2025FTT, ShestakovKorshunov2025JSNM}. Its detailed derivation is given in~\cite{ShestakovKorshunov2025JSNM}, so here we write the resulting expression without repeating the derivation:
\begin{eqnarray}\label{eq:DeltaOmega2}
 &\Delta\Omega(T) = - \pi T\sum_{\omega_n}\sum_{\alpha = a,b} N_{\alpha}\left[ \frac{\omega_n\tomega_{\alpha n}}{\sqrt{\tomega_{\alpha n}^2 + \tphi_{\alpha n}^2}}\right. \\ &+ \left. \sqrt{\tomega_{\alpha n}^2 + \tphi_{\alpha n}^2} -\left|\omega_n\right| - \left|\tomega_{\alpha n}^{\mathrm{N}}\right| \right]
 + \Delta\tilde{\Omega}(T),\nonumber
\end{eqnarray}
where $N_\alpha$ is a density of states at the Fermi level in the normal phase within a band having index $\alpha = (a,b)$, $\omega_n = 2(n + 1)\pi T$ is the fermionic Matsubara frequency, $\tomega_{\alpha n}$ and $\tphi_{\alpha n}$ are Matsubara frequency and order parameter, respectively, renormalized by the superconducting interaction and nonmagnetic impurity scattering, with $\tomega_{\alpha n}^{\mathrm{N}}$ is the renormalized Matsubara frequency at the normal state.

Impurity concentration is given as $n_\imp$, and impurity scattering potential is divided into an intraband, $v$, and interband, $u$, parts relating to each other as $\eta = v/u$. In the Born limit, when $\pi u N_\alpha \ll 1$ we have $\Delta\tilde{\Omega}=0$, whereas in the general case 
\begin{eqnarray}\label{eq:DeltaOmega2p}
 \Delta\tilde{\Omega}(T) &= \pi T N_a\Gamma_a\sum_{\omega_n}\left[ \frac{2\sigma(1-\eta^2)^2 + (1-\sigma)\kappa_{\imp}}{2\Den_{\imp}}\right. \nonumber\\ &- \left. \frac{2\sigma(1-\eta^2)^2 + (1-\sigma)\kappa_{\imp}^{\mathrm{N}}}{2\Den_{\imp}^{\mathrm{N}}} \right] \\ &- n_{\imp}T\sum_{\omega_n}\ln{\left(
 \Den_{\imp}/\Den_{\imp}^\mathrm{N} 
 \right)},\nonumber
\end{eqnarray}
where 
\begin{equation}\label{eq:kappa.imp}
 \kappa_{\imp} = \eta^2\frac{N_a^2 + N_b^2}{N_aN_b} + 2\frac{\tomega_{an}\tomega_{bn} + \tphi_{an}\tphi_{bn}}{\sqrt{\tomega_{an}^2 + \tphi_{an}^2}\sqrt{\tomega_{bn}^2 + \tphi_{bn}^2}},
\end{equation}
\begin{equation}\label{eq:D_imp}
 \Den_{\imp} = (1-\sigma)^2 + \sigma^2(1-\eta^2)^2 + \sigma(1-\sigma)\kappa_{\imp},
\end{equation}
$\kappa_{\imp}^{\mathrm{N}} = \Bigl.\kappa_{\imp}\Bigr|_{\tphi_{\alpha n} = 0}$, $\Den_{\imp}^{\mathrm{N}} = \Bigl.\Den_{\imp}\Bigr|_{\tphi_{\alpha n} = 0}$. Here a pair of parameters $\sigma$ and $\Gamma_a$, are introduced. The first one is a generalized scattering cross-section, 
\begin{equation}\label{eq:sigma}
 \sigma = \frac{\pi^2N_aN_bu^2}{1 + \pi^2N_aN_bu^2}.
\end{equation}
It represents strength of the scattering potential of the impurity and can vary from $0$ for the weak impurity scattering in the Born limit ($\pi u N_\alpha \ll 1$) to $1$ in the unitary limit for the strong scattering impurity ($\pi u N_\alpha \gg 1$). The second parameter is the impurity scattering rate, 
\begin{equation}\label{eq:Gamma_a}
 \Gamma_a = \frac{2n_{\imp}\sigma}{\pi N_a} = 2n_{\imp}\pi N_b u^2(1-\sigma).
\end{equation}
The parameters $\sigma$ and $\Gamma_a$ contain only the interband part of the impurity potential, $u$, and, therefore, do not depend on relation  $\eta$. It should be noted that $\Gamma_a$ depends on impurity concentration, $n_{\imp}$, which makes it convenient to be used for estimating the effect of disorder in experiments where the change in concentration of defects within samples is caused, for instance, by proton irradiation~\cite{Ghigo2018}.

Variables $\tomega_{\alpha n}$ and $\tphi_{\alpha n}$ are self-consistently calculated by solving the Eliashberg equations: 
\begin{eqnarray}
 &\ii\tomega_{\alpha n} = \ii\omega_n - \Sigma_{0}^{\SC}(\alpha, n) - \Sigma_{0}^{\imp}(\alpha, n),\label{eq:RenormOmega}\\
 &\tphi_{\alpha n} = \Sigma_{2}^{\SC}(\alpha, n) + \Sigma_{2}^{\imp}(\alpha, n),\label{eq:RenormPhi}
\end{eqnarray}
where the self-energy $\Sigma^{\SC}$ is defined by the coupling interaction and depends on a $2 \times 2$ coupling constant matrix having the elements $\lambda_{\alpha\beta}$ within the band space, with the self-energy $\Sigma^{\imp}$ corresponds to the nonmagnetic impurity scattering and it is obtained within the $\mathcal{T}$-matrix approximation that is equivalent to the non-crossing diagram approximation. Indices ``$0$'' and ``$2$'' denote the corresponding Pauli matrices $\hat{\tau}_i$ within the Nambu space. Within the system of units employed here with $\hbar = \kB = 1$, temperature $T$ and frequency $\omega_n$ are given in units of energy.

\section{Results of calculations.}\label{txt:results}
\begin{figure}
	\centering
	(a)\includegraphics[width=0.7\linewidth]{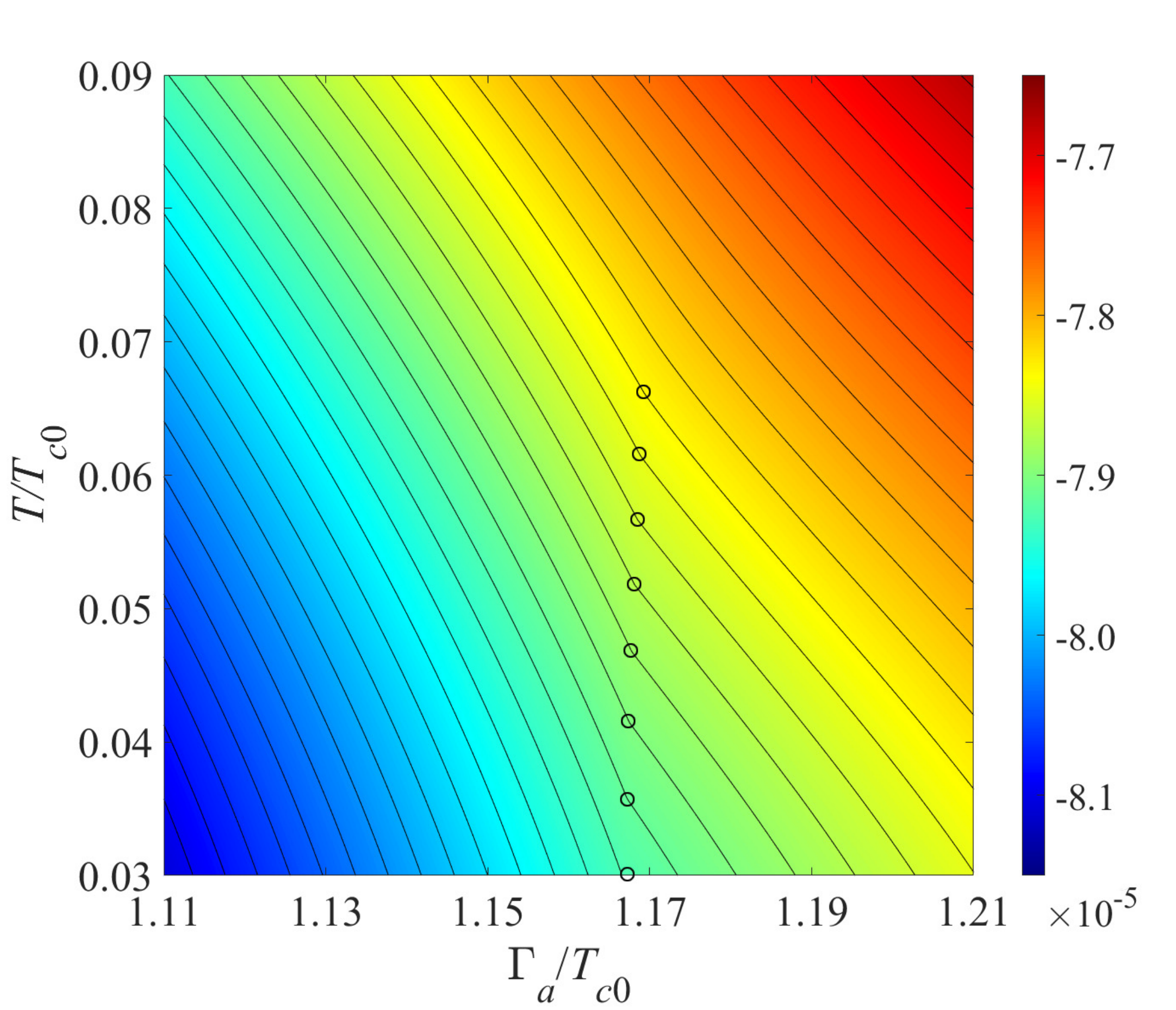} \\
	(b)\includegraphics[width=0.7\linewidth]{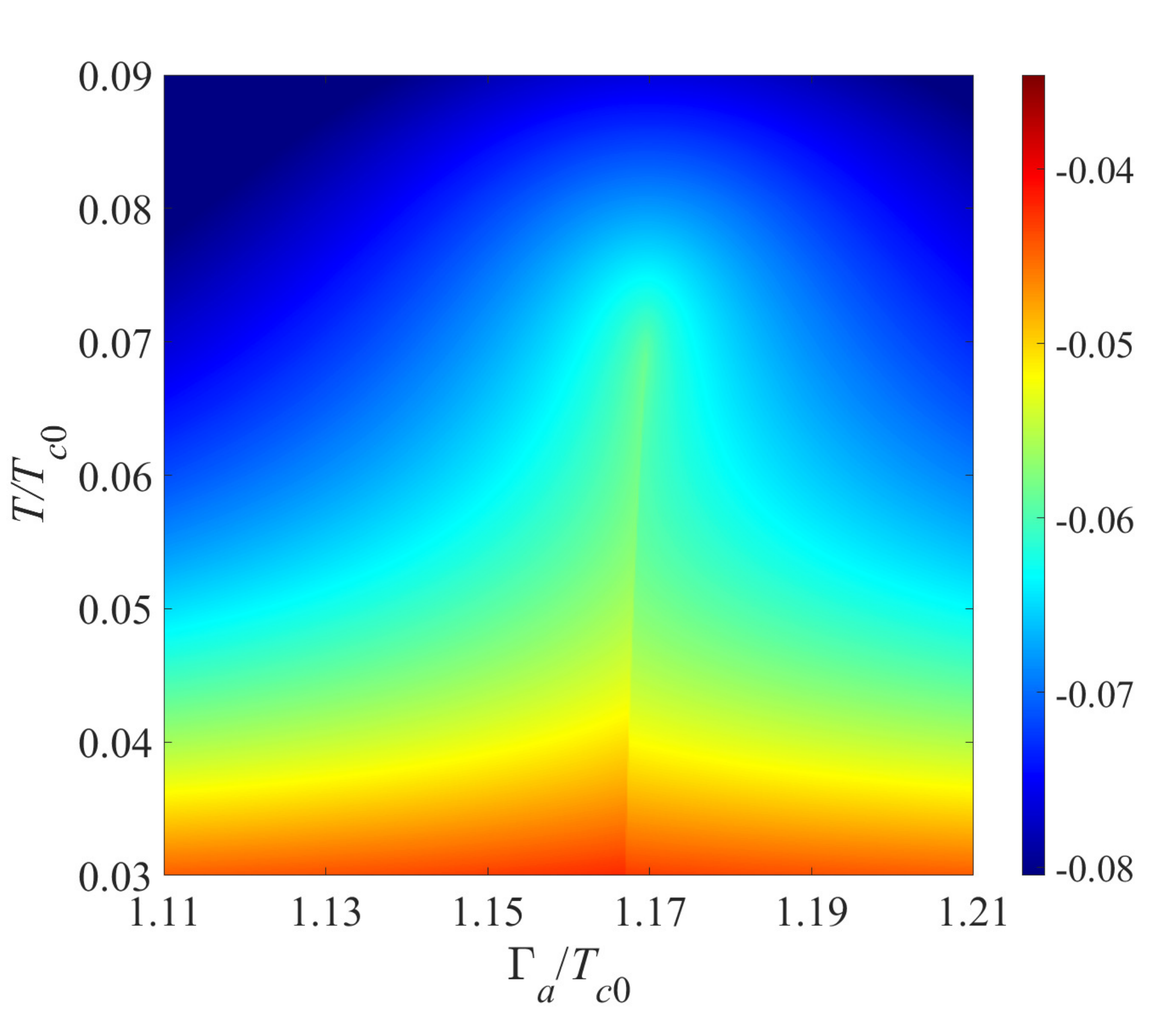} \\
	(c)\includegraphics[width=0.7\linewidth]{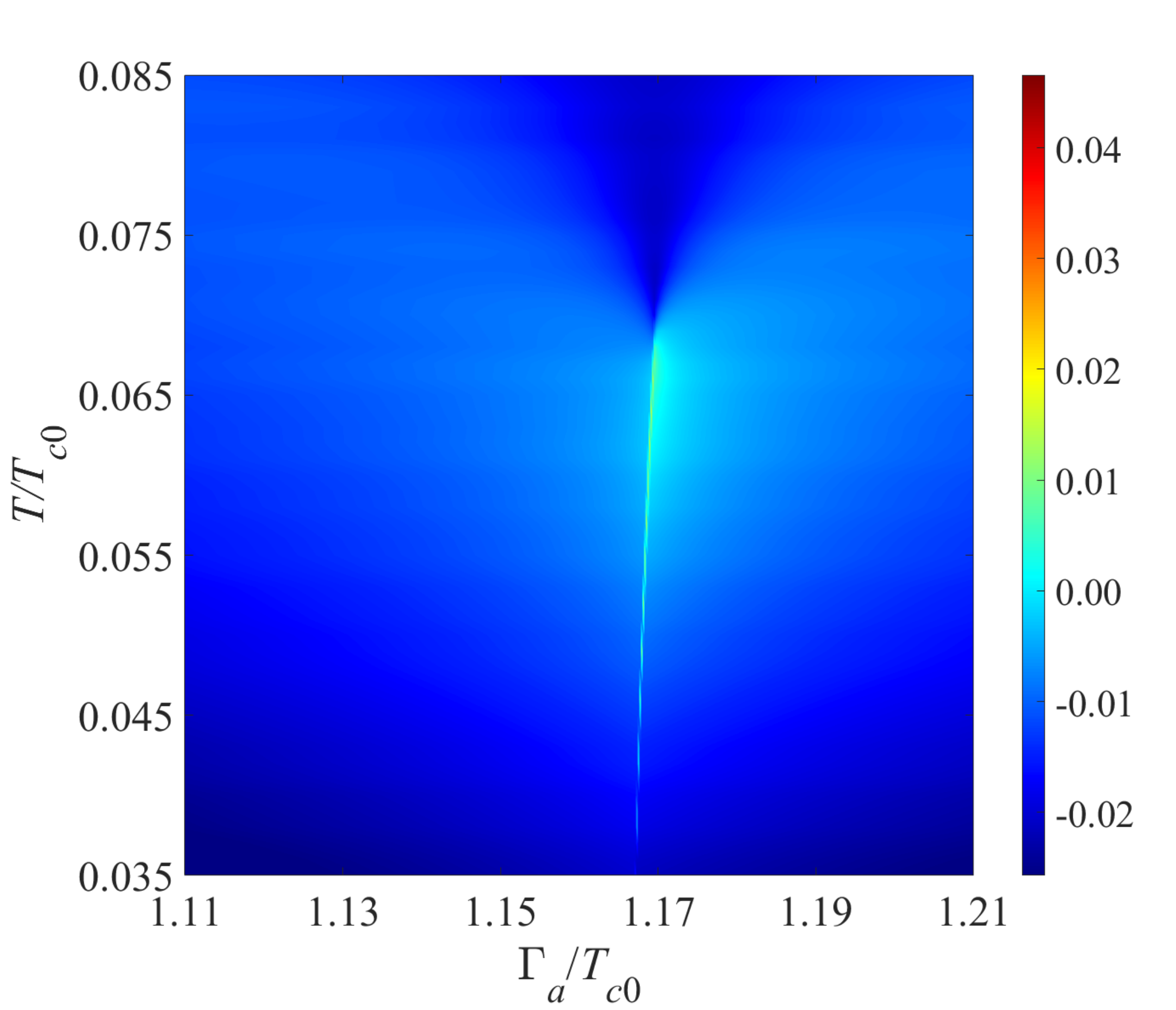} 
	\caption{(Color online) The surface of the minimum energy $\Delta\Omega(\Gamma_a, T)$ (a) in the Born limit, $\sigma = 0$, has the kink within the region of coexistence of two solutions of the Eliashberg equations. For clarity, constant level lines are drawn on the surface, with the kink being denoted with circle markers. The presence of the kink in $\Delta\Omega$ leads to the line of peaks in entropy $\Delta{S(\Gamma_a, T)}$ (b) along with line of divergencies in electronic specific heat $\Delta{C(\Gamma_a, T)}/T$ (c). Free energy, entropy and specific heat are given in units of eV, J/(K$\cdot$mol) and J/(K$^2\cdot$mol), respectively}\label{fig:Deltas}
\end{figure}
\begin{figure}
	\centering
	(a)\includegraphics[width=0.7\linewidth]{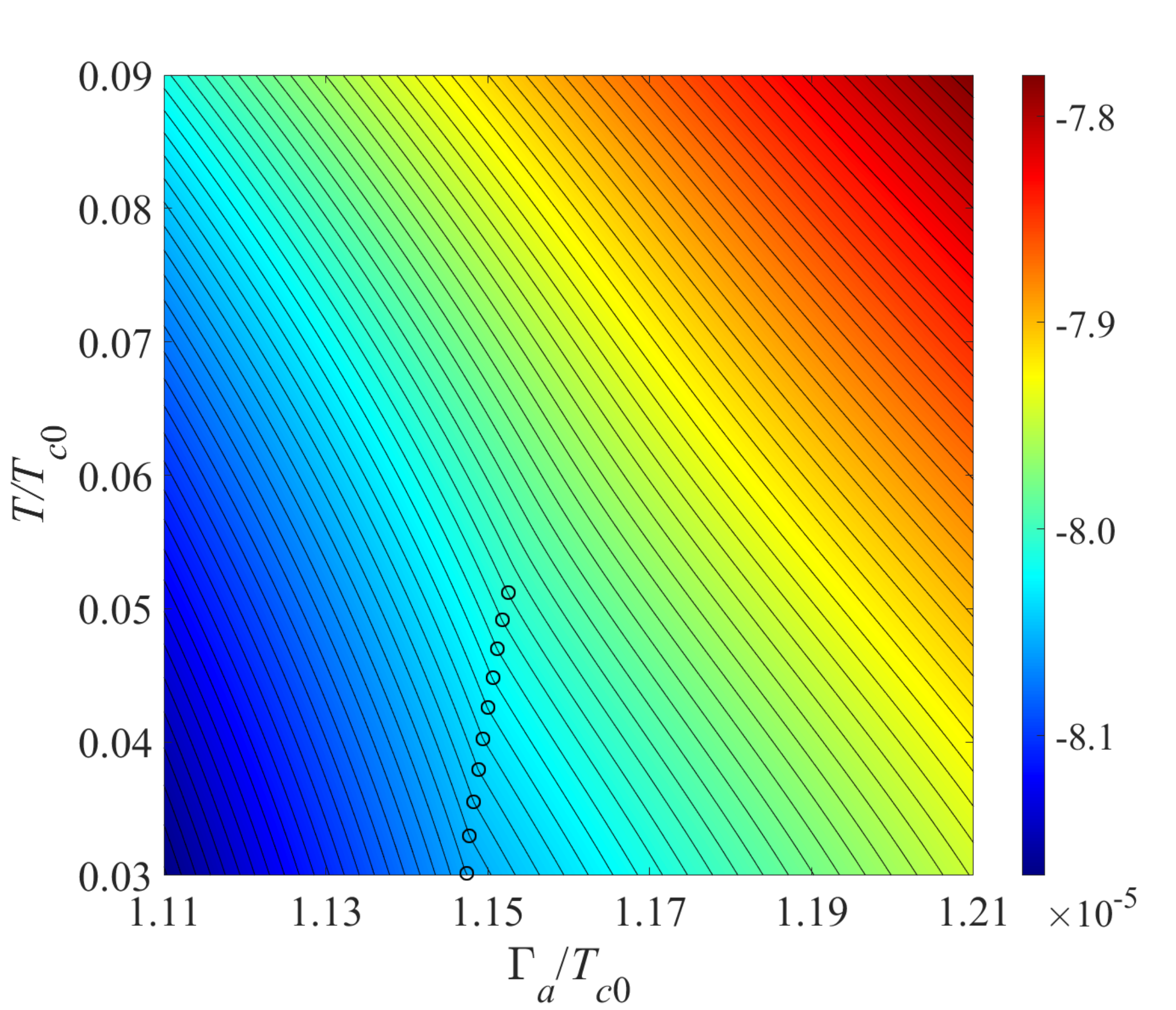} \\
	(b)\includegraphics[width=0.7\linewidth]{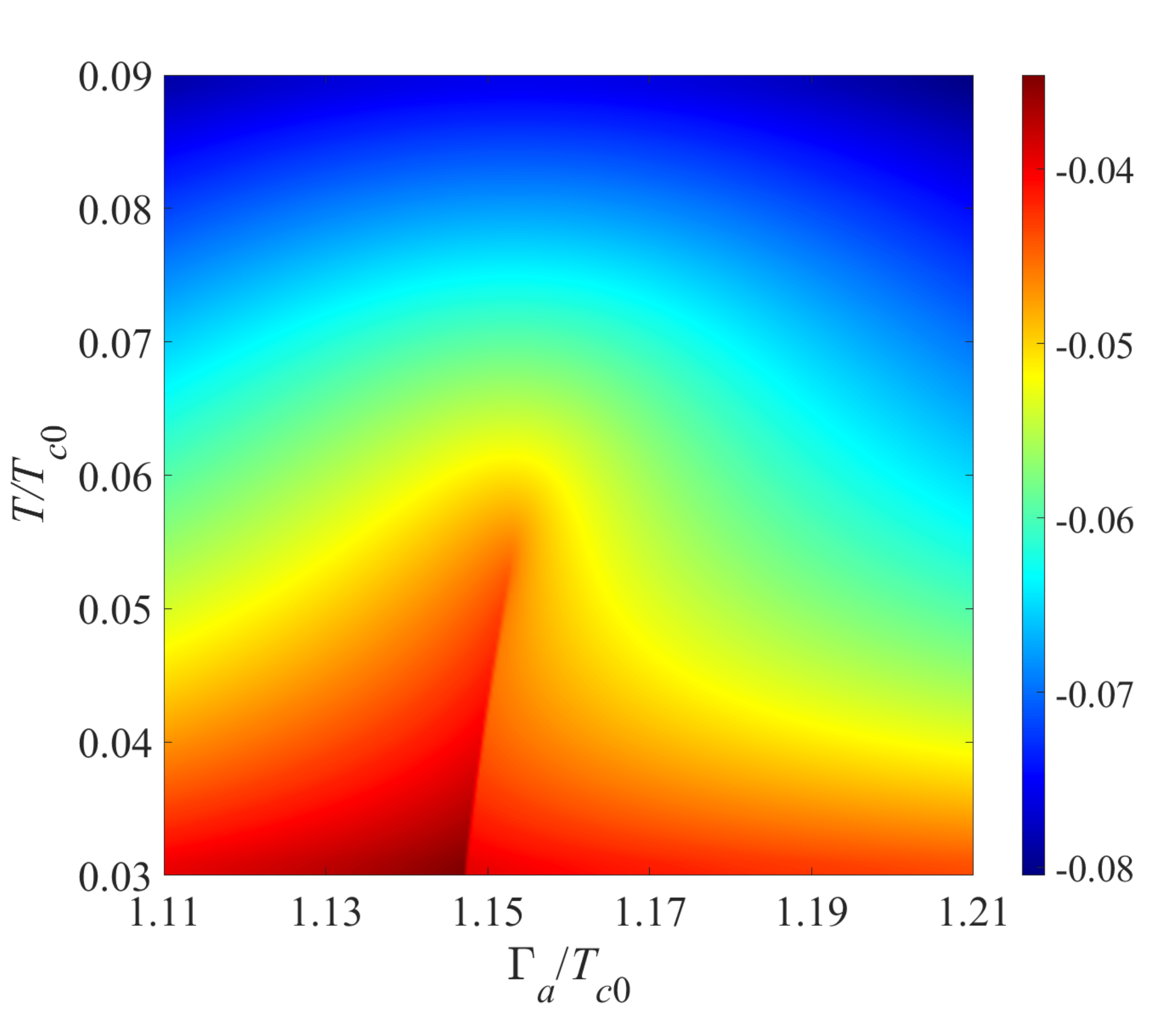} \\
	(c)\includegraphics[width=0.7\linewidth]{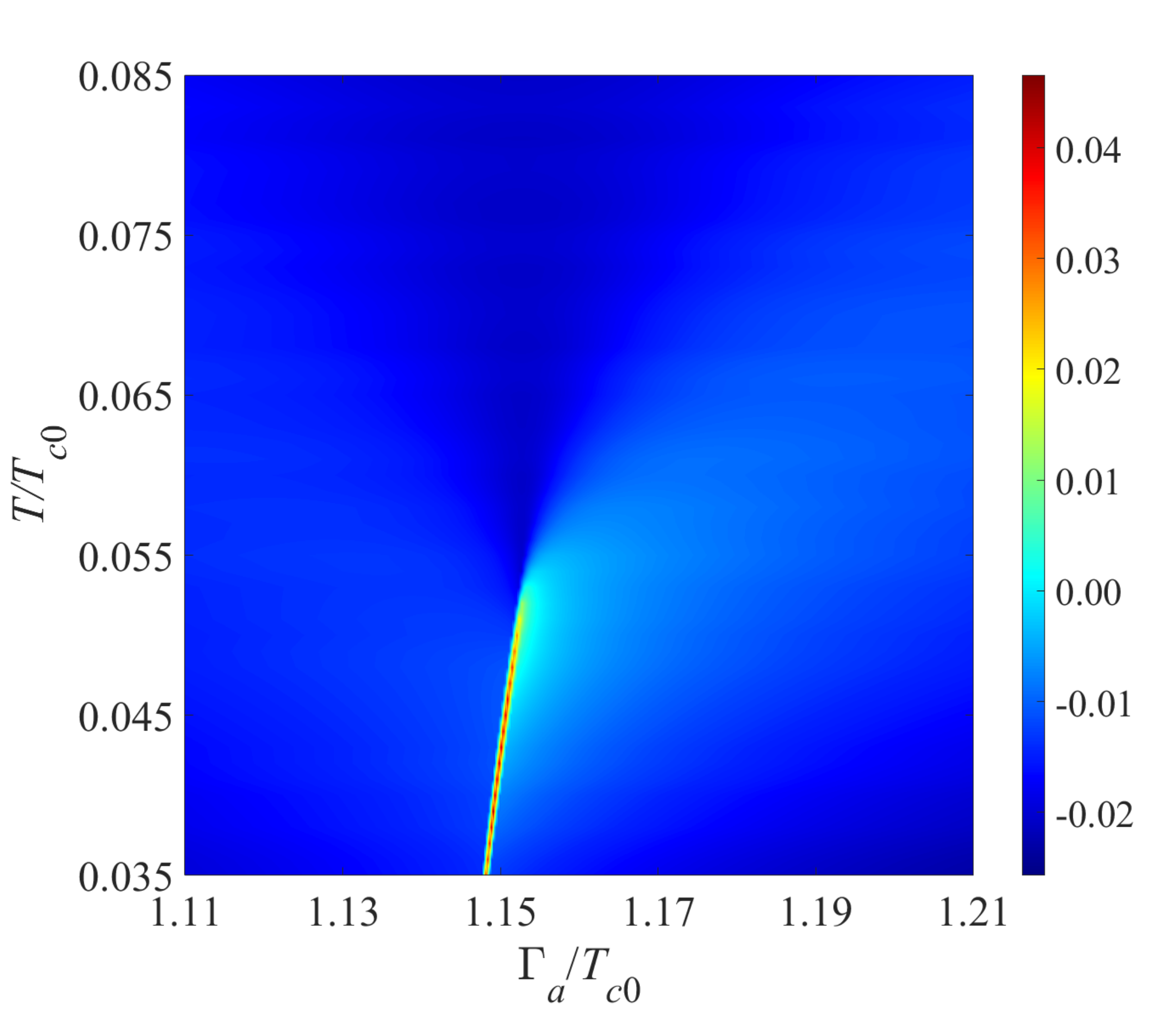}
	\caption{(Color online) Surfaces of $\Delta\Omega$ (a), $\Delta S$ (b) and $\Delta C/T$ (c) in the axes ($\Gamma_a$, $T$) near the Born limit at $\sigma = 0.06$. The main difference from the results shown in Fig.~\ref{fig:Deltas} is the decrease of the critical end point temperature $T_{\CEP}$. All values are given in the same units as in Fig.~\ref{fig:Deltas}}\label{fig:Deltas2}
\end{figure}
The impurity scattering induced transition between $s_{\pm}$ and $s_{++}$ states takes place if the coupling constant averaged over bands, $\langle \lambda \rangle = \left[\left(\lambda_{aa} + \lambda_{ab}\right) N_a + \left(\lambda_{ba} + \lambda_{bb}\right) N_b\right]/\left(N_a + N_b\right)$, of the clean $s_{\pm}$ superconductor is positive. With this in mind, in the following calculations we set the coupling constant matrix elements to be: $\lbrace \lambda_{aa}, \lambda_{ab}, \lambda_{ba}, \lambda_{bb} \rbrace = \lbrace 3.0, -0.2, -0.1, 0.5 \rbrace$. In the clean limit, the values lead to superconductivity with the critical temperature $T_{c0} = 40$~K. Densities of states  $N_a = 1.0656$~eV$^{-1}$ and $N_b = 2N_a$ are chosen for the full density of states, $N = N_a + N_b$, to match the one obtained from calculations within the density functional theory~\cite{Ferber2010, Yin2011, Sadovskii2012}. Without loss of generality, we assume that the impurity scattering happens within the interband channel only, $\eta = 0$. It was previously shown~\cite{ShestakovKorshunovSUST2018} that the finite value of the intraband part of the impurity scattering potential does not affect the superconducting state in the Born limit, while for finite values of $\sigma$, it only shifts the point of the transition to higher values of $\Gamma_a$.

Results of the calculations in the Born limit with $\sigma = 0$ and in its vicinity with $\sigma = 0.06$ are shown in Figs.~\ref{fig:Deltas} and~\ref{fig:Deltas2}. The impurity scattering rate in this case is defined by the expression after the second equal sign in equation~(\ref{eq:Gamma_a}).

In Fig.~\ref{fig:Deltas}~(a), a surface of the minimum energy $\Delta\Omega(\Gamma_a, T)$ is shown in the Born limit. Within the temperature range $T < 0.07$~$T_{c0}$ and range of impurity scattering rates $1.16 < \Gamma_a/T_{c0} < 1.17$, there is a kink on the surface of $\Delta\Omega$. This kink corresponds to the line of the $s_{\pm} \to s_{++}$ transition at which a sign of a gap function $\Delta_b$ changes abruptly. Such a behavior of the superconducting gap along with the presence of the kink at the free energy surface, and existence of two sets of solutions of the Eliashberg equations within considered regions points out that the nonmagnetic impurity scattering induced transition between $s_{\pm}$ and $s_{++}$ states in the Born limit is the first order phase transition. At temperatures $T > 0.07$~$T_{c0}$, there is no any peculiarity in the free energy, the Eliashberg equations have unique solutions, and smooth sign changing in $\Delta_b$ corresponds to the smooth transition, i.~e. a crossover, between the $s_{\pm}$ and $s_{++}$ states. The transformation of the first order phase transition to the crossover occurs at the critical end point (CEP) at temperature $T_{\CEP} \approx 0.07$~$T_{c0}$. At this temperature, the range of $\Gamma_a$ values at which different solutions of the Eliashberg equations coexist reduced to a single point $\Gamma_a^{\CEP}$.

\begin{figure}
	\centering
	(a)\includegraphics[width=1.0\linewidth]{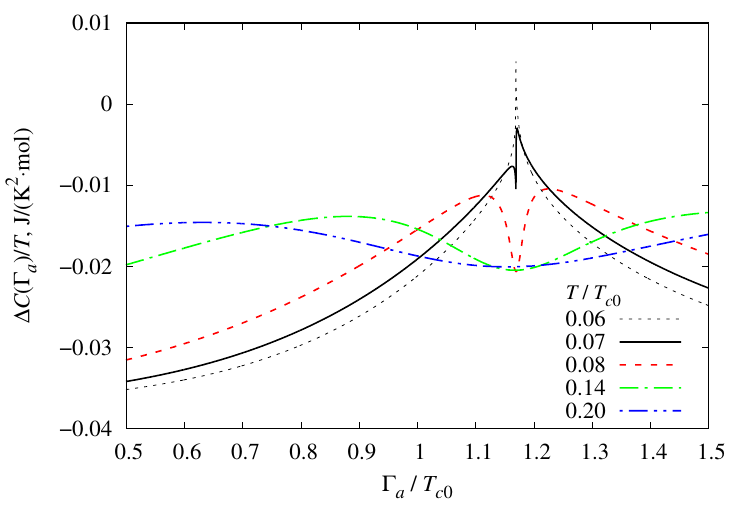} \\
	(b)\includegraphics[width=1.0\linewidth]{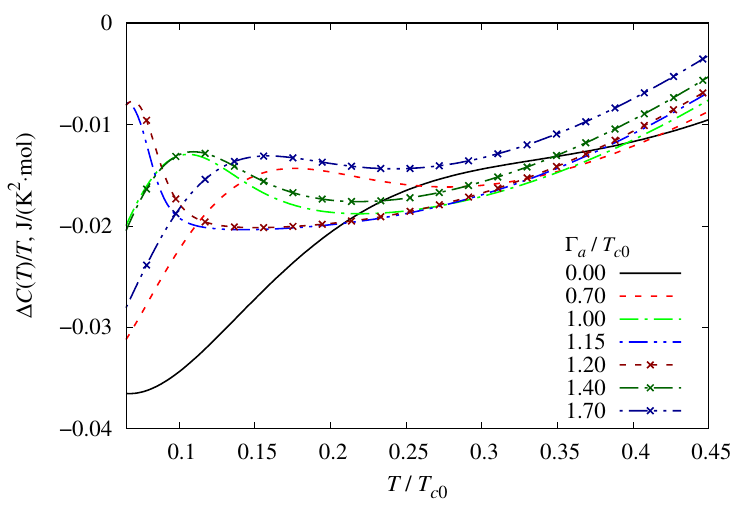} 
	\caption{(Color online) Graphs for $\Delta{C}(\Gamma_a)/T$ (a) and $\Delta{C}(T)/T$ (b) plotted for the temperature above the critical end point $T_{\CEP}$ in the Born limit. The panel (a) shows also plot for $\Delta{C}(\Gamma_a)/T$ at $T = 0.06$~$T_{c0}$ revealing how specific heat diverges due to the first order phase transition. In panel (b), marked and unmarked lines correspond to the families of curves located to the left and to the right of the critical end point $\Gamma_a^{\CEP}$, respectively}\label{fig:DeltaC}
\end{figure}

The presence of the kink in $\Delta\Omega(\Gamma_a, T)$ leads to peculiarities in its first and second derivatives with respect to temperature: a peak in the change of entropy, 
\begin{equation}\label{eq:entropy}
	\Delta S(T) = -\frac{\partial\Delta\Omega(T)}{\partial T},
\end{equation}   
and a divergence in the change of electronic specific heat,
\begin{equation}\label{eq:specific.heat}
	\Delta C(T) = -T\frac{\partial^2\Delta\Omega(T)}{\partial T^2}.
\end{equation}
These quantities are presented in Figs.~\ref{fig:Deltas}~(b) and (c). A detailed description of the aforementioned peculiarities are given in the work~\cite{ShestakovKorshunovSUST2025}, therefore, we only focus on the following important points. Firstly, note that the first order phase transition line is most clearly visible in the plot of ~$\Delta{C}(\Gamma_a, T)$. Secondly, we consider how electronic specific heat depends on temperature and concentration of disorder that can be important for experimental observation of the results presented. Above the critical end point temperature $T_{\CEP}$, the single peculiarity in $\Delta{C}(\Gamma_a)$ caused by the abrupt sign change in the gap function $\Delta_b$ splits out to a pair of the peaks, see Fig.~\ref{fig:DeltaC}~(a). The value of $\Gamma_a$ corresponding to $\Delta_b$ going through zero while changing its sign is located between these peaks. Increasing the temperature smears each peak, with the distance between them being increased. While the amount of disorder grows, the peak in the temperature dependence of $\Delta{C}(T)$ presented in Fig.~\ref{fig:DeltaC}~(b) becomes more prominent, with its maximum shifting to the low temperatures. When the extent of disorder exceeds $\Gamma_a^{\CEP}$, the peak behaves in the opposite manner: further increase of disorder smears the peak and shifts its maximum to higher temperatures.

\begin{figure}
	\centering
	\includegraphics[width=1.0\linewidth]{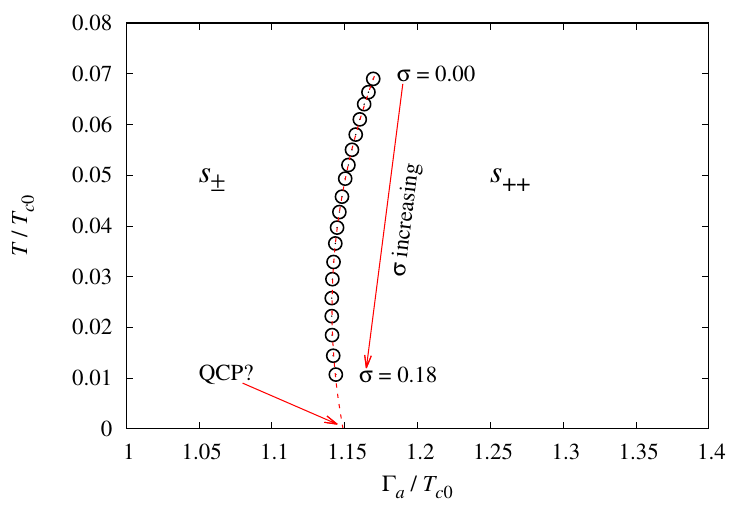}
	\caption{(Color online) Location of the critical end point at the ($\Gamma_a$, $T$) plane for different values of the generalized cross-section $0.0 < \sigma < 0.18$. The dashed line corresponds to the critical end point location extrapolated to the zero temperature, where it may become the quantum critical point at $\Gamma_a^{\mathrm{QCP}}\approx 1.15$~$T_{c0}$}\label{fig:CEP}
\end{figure}

Results near the Born limit with $\sigma = 0.06$ are similar to those presented above. Graphs for $\Delta\Omega$, $\Delta S$ and $\Delta C$ are plotted in Fig.~\ref{fig:Deltas2}. The main difference from the Born limit with $\sigma = 0$ is the lowering of the temperature $T_{\CEP}$ and decrease of $\Gamma_a^{\CEP}$. The shift of critical end point position with the increasing $\sigma$ can be traced on the ($\Gamma_a, T$) plane, see Fig.~\ref{fig:CEP} where the critical end point is depicted with circle markers. The temperature $T_{\CEP}$ follows a monotonic rule: ``higher the parameter  $\sigma$, lower the temperature $T_{\CEP}$'', whereas $\Gamma_a^{\CEP}$ is initially decreases for $0 <\sigma < 0.15$ and then increases. Previously, it was shown that for the cross-section $\sigma > 0.12$ and temperature $T \geq 0.03T_{c0}$ the abrupt transition between $s_{\pm}$ and $s_{++}$ states is absent~\cite{ShestakovKorshunovSUST2018, ShestakovKorshunovSymmetry2018, ShestakovKorshunovSUST2025}. Here we show that the abrupt transition appears to exist even for lower temperatures until $T = 0.01$~$T_{c0}$. Since calculations are performed for discrete Matsubara frequencies, in order to trace the position of the critical end point to even lower temperatures, we made an extrapolation to the zero temperature, see dashed line in Fig.~\ref{fig:CEP}. The best fitting function is found to be a second-order polynomial.
Crossing the dashed line with the line of $T = 0$ points out to the possibility for the critical end point with increasing $\sigma$ to become a quantum critical point (QCP) at zero temperature and impurity scattering rate $\Gamma_a^{\mathrm{QCP}}\approx 1.15$~$T_{c0}$.

\section{Conclusions.}\label{txt:concl}
For iron-based superconductors within the two-band model with the nonmagnetic impurity scattering treated in the approximation of non-crossing diagram ($\mathcal{T}$-matrix approximation), the change in the Grand thermodynamic potential (the Landau free energy) at the transition into the superconducting state $\Delta\Omega = \Omega_{\mathrm{S}} - \Omega_{\mathrm{N}}$ is calculated. In the Born limit with $\sigma = 0$, in the parameter space ($\Gamma_a$, $T$) there is the line of the first order phase transition induced by nonmagnetic impurity scattering accompanied by the abrupt change of the superconducting order parameter structure $s_{\pm} \leftrightarrow s_{++}$. The line of the phase transition is limited from above by the critical end point at the temperature $T_{\mathrm{CEP}}$ and impurity scattering rate $\Gamma_a^{\mathrm{CEP}}$. With increasing the strength of the impurity scattering potential (increasing $\sigma$), the temperature $T_{\mathrm{CEP}}$ is decreasing down to $T_{\mathrm{CEP}} = 0.01 T_{c0}$ at $\sigma = 0.18$. Previously, we shown that the abrupt transition is smeared out at $\sigma \approx 0.12$ and temperature $T = 0.03$~$T_{c0}$~\cite{ShestakovKorshunovSUST2018, ShestakovKorshunovSymmetry2018, ShestakovKorshunovSUST2021, ShestakovKorshunovSUST2025, ShestakovKorshunov2025FTT}. The result obtained here is in agreement with the previous findings and refines them. The extrapolation of the critical end point to zero temperature shows that with increasing $\sigma$ it may become the quantum critical point at $T = 0$, $\Gamma_a^{\mathrm{QCP}}\approx 1.15 T_{c0}$ and $\sigma^{\mathrm{QCP}} \approx 0.21$. Similar behavior is observed at metamagnetic transition in bilayer samples of the strontium ruthenate, Sr$_3$Ru$_2$O$_7$,~\cite{Grigera2001} or rare metal titanates~\cite{Wang2022}. Nevertheless, the confirmation of the existence of the quantum phase transition demands additional studies at zero temperature.

\textbf{Acknowledgments.}\label{txt:ackn}
We are grateful to A. S. Mel'nikov and A. K. Murtazaev for helpful discussions.

\textbf{Funding.}\label{txt:fin}
The study was supported by the Russian Science Foundation grant no.~25-22-20043, \href{https://rscf.ru/project/25-22-20043/}{https://rscf.ru/project/25-22-20043/}, grant of the Krasnoyarsk Regional Science.

\textbf{Conflict of interests.} The authors of this work declare that they have no conflicts
of interest.

\end{document}